\documentclass[
    ,final            
  ]
  {aipproc}

\layoutstyle{6x9}

\begin{document}

\title{Probing cosmological parameters with GRBs}

\author{T. Di Girolamo}{
  address={INFN, Sezione di Napoli, Italy}
}

\author{M. Vietri}{
  address={Scuola Normale Superiore, Pisa, Italy}
}

\author{G. Di Sciascio}{
  address={INFN, Sezione di Napoli, Italy}
}

\begin{abstract}
In light of the recent finding of the narrow clustering of the 
geometrically-corrected gamma-ray energies emitted by Gamma Ray Bursts (GRBs),
we investigate the possibility to use these sources as standard candles
to probe cosmological parameters such as the matter density $\Omega_m $ and
the cosmological constant energy density $\Omega_{\Lambda}$. By simulating
different samples of gamma-ray bursts, based on recent observational results,
we find that $\Omega_m $ (with the prior $\Omega_m + \Omega_{\Lambda} = 1$) 
can be determined with accuracy $\sim$7\% with data from 300 GRBs, 
provided a local calibration of the standard candles be achieved.   
\end{abstract}

\maketitle


\section{Introduction}

Recent studies have pointed out that Gamma-Ray Bursts (GRBs) may be
considered as standard cosmological candles. 
The prompt $\gamma$-ray energies of GRBs, after correction for the
conical geometry of the jet, result clustered around a mean value of
a few 10$^{50}$ erg \cite{Frail}. 

Since the discovery that GRBs lie at cosmological
distances, about 30 redshifts have been measured. Apart from the
controversial case of GRB 980425, possibly associated with the nearby
supernova SN1998bw (at $z=0.0085$),
all other redshifts are spread within the wide 0.17$-$4.5 range. 
Therefore GRBs could be good candles to probe cosmological parameters
\cite{Schaefer} \cite{Japan}.

GRBs are thought to be associated with the death of
massive (and short lived) stellar progenitors.
Therefore the rate of GRB events per unit cosmological volume should be a
tracer of the global history of star formation. 

Hence, we have now all the
information necessary to perform simulations of GRB distributions in a
given cosmological model. Universal parameters such as the matter density
fraction $\Omega_m$ and the cosmological constant energy
fraction $\Omega_{\Lambda}$,
can be constrained by fitting the Hubble diagrams corresponding to such
simulated distributions. It is the aim of this paper to simulate
different GRB distributions and investigate their ability to determine
the cosmological parameters $\Omega_m$ and $\Omega_{\Lambda}$. Both universes
with and without a cosmological constant $\Lambda$ will be considered.

\section{To start: a KS test}

\begin{figure}
  \includegraphics[height=.3\textheight,angle=270]{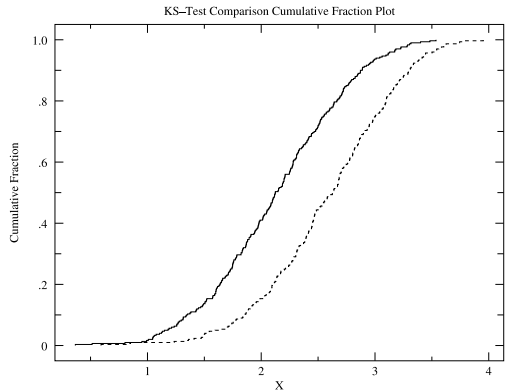}
  \caption{Comparison of the cumulative fractions obtained with the parameter
$X\equiv \log d_L^2 (Gpc)$ calculated for two data sets of 300 GRBs simulated
in an universe with $\Omega_m = 1$ and $\Omega_{\Lambda} = 0$ (full line)
and in one with $\Omega_m = 0.3$ and $\Omega_{\Lambda} = 0.7$ (dashed line).
The corresponding probability that the two data sets are drawn from the
same distribution is $Q_{KS} = 2.48\cdot 10^{-14}$.}
\label{KS}
\end{figure}

First, in order to show what we are 
aiming at, we performed a Kolmogorov-Smirnov (KS) test on two data sets 
made of 300 GRBs simulated in two different cosmological models, one with
$\Omega_m = 1$ and $\Omega_{\Lambda} = 0$ and the other with 
$\Omega_m = 0.3$ and $\Omega_{\Lambda} = 0.7$, but both with 
a Hubble constant $H_0 = 65$ km s$^{-1}$ Mpc$^{-1}$ (as it will be 
assumed throughout the paper). 
We assume that GRBs are indeed standard candles with true prompt
$\gamma$-ray energy released, $E_{\gamma} $, following a Gaussian distribution 
in its logarithm with mean $\mu = 50.7$ (if $E_{\gamma} $ is expressed in 
$erg$ units) and $\sigma =0.3$ (corresponding to a multiplicative factor of 2) 
\cite{Frail}, and that they are distributed in the universe
according to the model of star formation rate
$R_{SF1} (z)$ reported in \cite{Porciani}, which matches the 
$\log N - \log P$ relation (GRB number counts vs. peak photon
flux) obtained with BATSE data. Applying the KS test on the redshift
distributions, we found that the probability that the two data sets are
drawn from the same distributions
is $Q_{KS}=0.031$, a ``no man's land'' value for this test. On the
other hand, the application of the KS test on the parameter 
$\log d_L^2 (z)$, where $d_L (z)$ is the luminosity distance, resulted in a
significant probability $Q_{KS} \sim 10^{-14}$, which tells us that it is 
possible to discriminate between the two different cosmological models if a
set of 300 GRB luminosity distances is known (see Figure \ref{KS}).

\section{Data set simulations in a $\Lambda =0$ cosmology}

We consider now a $\Lambda =0$ 
cosmology, in which the only contribution to the density parameter is given 
by $\Omega_m$. We assume for GRBs the same energy distribution as for the KS
test. However, the assumed mean value is not relevant for our
investigation, since it is the dispersion value that constrains the
cosmological density parameter.  

The standard candle energy is related to the fluence of the 
burst $f_{\gamma} = E_{\gamma} (1+z)/(4\pi d_L^2 (z))$
via the luminosity distance $d_L (z)$.
In order to have a linear propagation of errors throughout our 
simulations, we choose to construct with GRBs a Hubble diagram 
$\log d_L^2 - z$, since the distribution of the parameter $\log d_L^2 $ is 
the same of that of $\log E_{\gamma}$, and therefore it is Gaussian. 

In order to study the ability of GRBs in probing the cosmological 
parameters as a function of their number, we have simulated different samples 
with $N_{GRB}$ = 10, 30, 100, 300 and 1000. Moreover, in order to be
free from statistical fluctuations, we have performed 10$^2$
realizations of each of these samples. 

The simulation of a GRB consists of the random sampling of both the 
redshift $z$ and the true $\gamma$-ray energy released $E_{\gamma} $,
according to the respective adopted distributions. Given a cosmological
model, from these coupled values we obtain the corresponding value for the 
parameter $\log d_L^2 $, which we plot on the Hubble diagram as a function 
of $z$. At this point we perform a $\chi^2 $ minimization of the simulated
data to see with which accuracy the fit reproduces the input cosmology.
The measurement error on $\log d_L^2 $ is assumed to be $\sigma = 0.3$. 
The mean results of our repeated fits in an Einstein-de Sitter universe 
($\Omega_m = 1$) are reported in the left side of Table~\ref{tablam}. 

\section{Data set simulations in a $\Lambda$-dominated cosmology}

We move now to a $\Lambda$-dominated cosmology, in which the  
contributions to the density parameter are given by the mass density, 
$\Omega_m$, and by the cosmological constant energy density, 
$\Omega_{\Lambda} $.
In light of the recent observations of the cosmic microwave background
anisotropy \cite{Bennett}, we put the prior of a flat universe 
$\Omega_m +\Omega_{\Lambda} =1$. 

\begin{table}[t]
\begin{minipage}[t]{.45\columnwidth}
\begin{tabular}{cccc}
\hline
    \tablehead{1}{r}{b}{$N_{GRB} $}
  & \tablehead{1}{r}{b}{$<\Omega_m >$}
  & \tablehead{1}{r}{b}{$<\Delta \Omega_m >$}
  & \tablehead{1}{r}{b}{$S_{\Omega_m}$}  \\
\hline
 10    & 0.9983 & 0.2997 & 0.3097 \\
 30    & 1.0158 & 0.1895 & 0.1993 \\ 
 100   & 0.9937 & 0.0993 & 0.1108 \\ 
 300   & 0.9959 & 0.0599 & 0.0629 \\
 1000  & 1.0009 & 0.0332 & 0.0351 \\
\hline
\end{tabular}
\end{minipage}
\begin{minipage}[t]{.5\columnwidth}
\begin{tabular}{ccccc}
\hline
    \tablehead{1}{r}{b}{$N_{GRB} $}
  & \tablehead{1}{r}{b}{$<\Omega_m >$}
  & \tablehead{1}{r}{b}{$<\Omega_{\Lambda} >$}
  & \tablehead{1}{r}{b}{$<\Delta \Omega >$}
  & \tablehead{1}{r}{b}{$S_{\Omega}$}  \\
\hline
 10    & 0.3195 & 0.6805 & 0.1004 & 0.1307 \\
 30    & 0.2973 & 0.7027 & 0.0763 & 0.0700 \\ 
 100   & 0.3002 & 0.6998 & 0.0363 & 0.0351 \\ 
 300   & 0.3023 & 0.6977 & 0.0219 & 0.0222 \\
 1000  & 0.3001 & 0.6999 & 0.0120 & 0.0125 \\
\hline
\end{tabular}
\end{minipage}
\caption{Mean values of the fitted cosmological density 
parameters $\Omega_m$ and $\Omega_{\Lambda} $, 
of their error $\Delta \Omega$ and their dispersion 
$S_{\Omega}$ obtained by fitting 10$^2$ GRB sample realizations 
with $N_{GRB} $ distributed according to function $R_{SF1}(z)$ of 
\cite{Porciani} in an Einstein-de Sitter universe ($\Omega_m = 1$, left)
and in a flat universe with input values 
$\Omega_m = 0.3$ and $\Omega_{\Lambda} = 0.7$ (right).}
\label{tablam}
\end{table}

Again, in order to study the ability of GRBs in probing the cosmological 
parameters in a $\Lambda$-dominated universe, we have simulated 10$^2$
realizations of GRB samples with $N_{GRB}$ = 10, 30, 100, 300 and 1000.
The $\chi^2 $ minimization of the resulting Hubble diagrams has been
performed considering $\log d_L^2 $ depending only on the fit parameter
$\Omega_m $, {\it i.e.}, using the relation $\Omega_{\Lambda} = 1 -\Omega_m $.
The right side of
Table \ref{tablam} reports the general results of our repeated fits 
for a flat cosmology with input values $\Omega_m = 0.3$ and 
$\Omega_{\Lambda} = 0.7$ (which are those adopted in \cite{Frail}).

\begin{figure}
\begin{minipage}[t]{.5\columnwidth}
\includegraphics[height=.3\textheight]{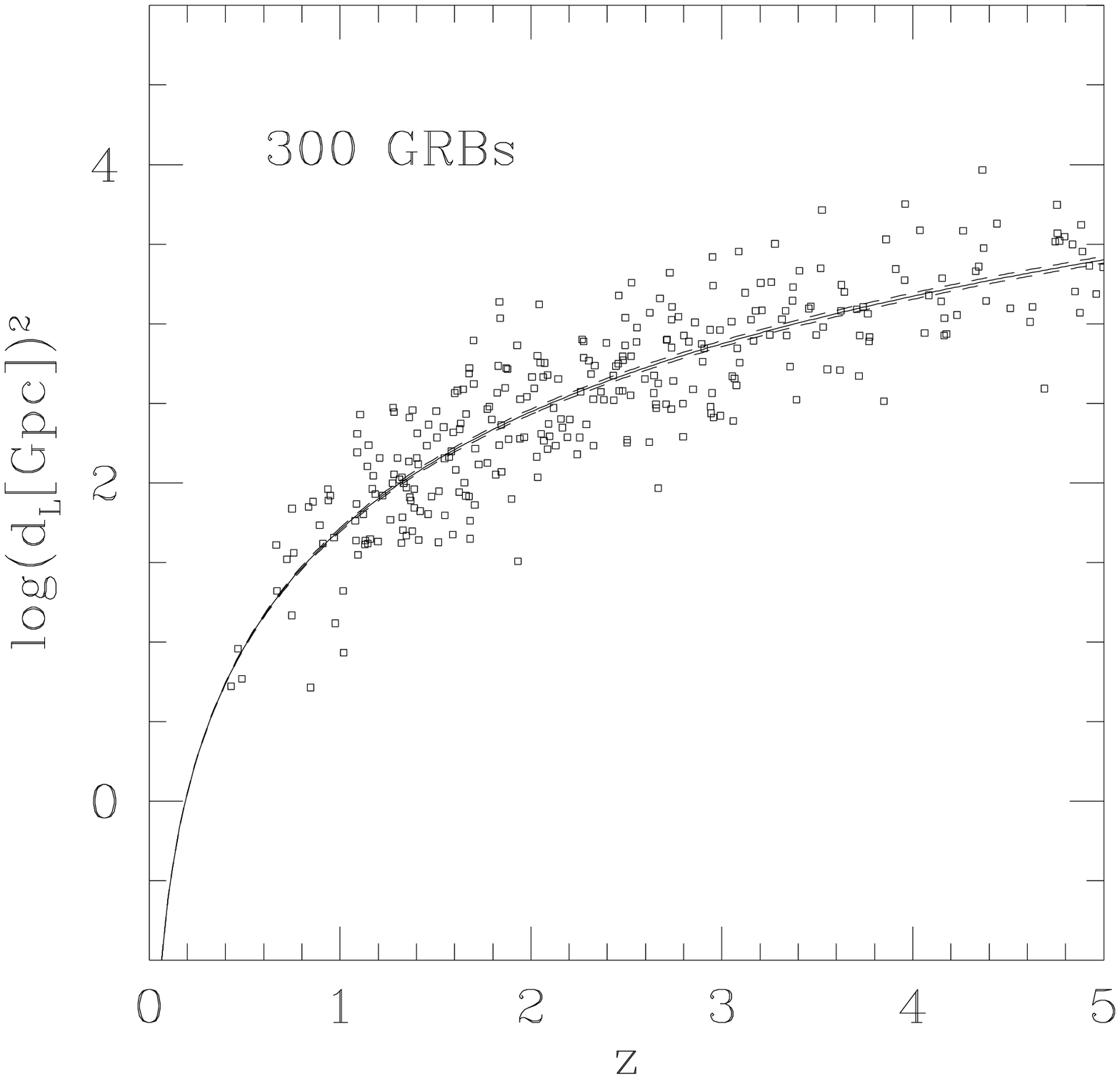}
\end{minipage}
\begin{minipage}[t]{.5\columnwidth}
\includegraphics[height=.3\textheight]{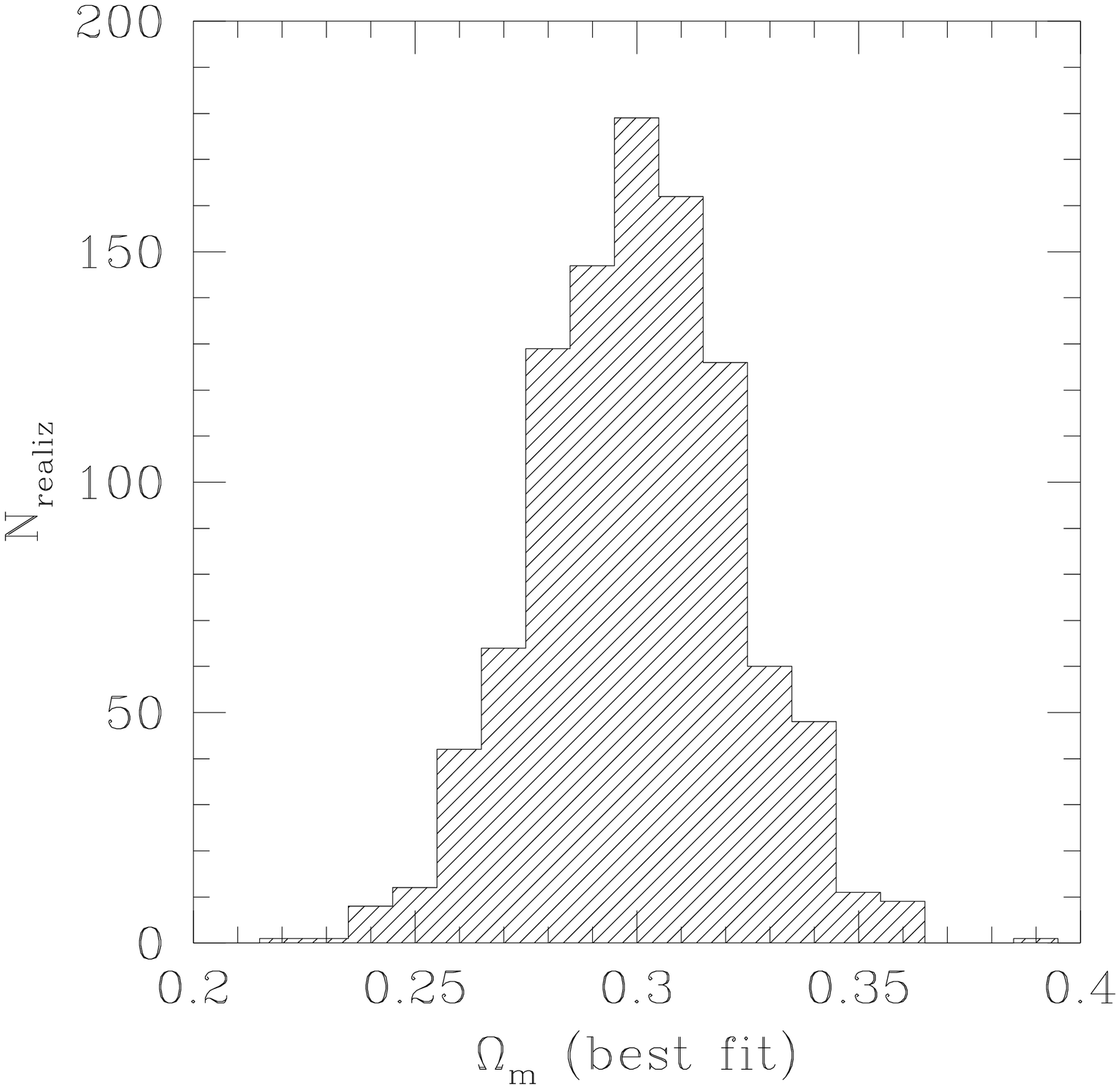}
\end{minipage}
\caption{Left: Hubble diagram $\log d_L^2 - z$ with data simulated
for a sample of 300 GRBs in a flat universe with density parameters
$\Omega_m = 0.3$ and $\Omega_{\Lambda} = 0.7$. The solid curve shows 
the function $\log d_L^2 (z)$ in the assumed cosmology, while the dashed
curves give the dispersion about the best fit parameter (upper curve
corresponds to lower $\Omega_m $). 
Right: Histogram with the distribution of the best fit
values of the matter density $\Omega_m $ for 10$^3$ realizations of
a sample of 300 GRBs in a flat universe with density parameters
$\Omega_m = 0.3$ and $\Omega_{\Lambda} = 0.7$. The distribution has
a mean $<\Omega_m > = 0.3001$, a median $\Omega_m (med) = 0.3002$,
a dispersion $S_{\Omega_m } = 0.0228$, and a kurtosis 
$k_{\Omega_m } = 3.0993$, to be compared with the value of a Gaussian
distribution, {\it i.e.}, 3.}
\label{hubblehisto}
\end{figure}

Focussing on the samples with 
$N_{GRB} = 300$, which represent the future data set expected from the
{\it Swift} satellite experiment, Figure \ref{hubblehisto} shows one of the 
Hubble diagrams $\log d_L^2 - z$ obtained with the simulations (left), together
with the distribution of the best fit values of the matter density fraction 
$\Omega_m $ for 10$^3$ sample realizations (right).  

Finally, we must remark that the analysis in \cite{Frail} 
assumes of course a particular set of 
cosmological parameters to derive the standard $\gamma$-ray 
energy of GRBs. To avoid a circular logic we should assume a candle 
calibration with a local sample of sources, a 
prospect which can now be considered possible in light of the discovery of 
the near GRB 030329, with redshift as low as $z=0.1685$.

\section{Conclusions}

We have simulated different samples of
GRBs adopting $\gamma$-ray energy and redshift distributions consistent
with recent observational results, in order to investigate their ability
to probe cosmological parameters such as the density fractions  
$\Omega_m $ and $\Omega_{\Lambda }$. Our result is that in a
$\Lambda $-dominated flat universe the accuracy
in the determination of the matter density $\Omega_m $ is  
$\sim$40\% for a sample with $N_{GRB} =10$ and an excellent $\sim$4\% for 
$N_{GRB} =1000$.

\end{document}